\documentclass[a4paper,UKenglish,cleveref, autoref, thm-restate]{lipics-v2021}

\pdfoutput=1 
\hideLIPIcs  

\nolinenumbers

\usepackage{amsfonts}
\usepackage{graphicx}
\usepackage{mathrsfs}
\usepackage{amssymb}
\usepackage{hyperref}
\usepackage{amsmath}
\usepackage{cite}
\usepackage{xcolor}
\usepackage[utf8]{inputenc}
\usepackage{mathtools,nicefrac}
\usepackage{amstext}
\usepackage{amsmath}
\usepackage{hyperref}
\usepackage{amsmath}
\usepackage{relsize}
\usepackage{amsmath,etoolbox}
\usepackage{algorithm2e}
\usepackage{algpseudocode}
\usepackage{amssymb}

\newcommand{\SW}{{\texttt{SW}}\xspace}
\newcommand{\USM}{{\texttt{USM}}\xspace}
\newcommand{\RRG}{{\texttt{RRG}}\xspace}

\DeclareMathOperator*{\argmax}{arg\,max}

\title{A Tight Competitive Ratio for Online Submodular Welfare Maximization}

\titlerunning{A Tight Competitive Ratio for Online Submodular Welfare Maximization}

\author{Amit Ganz}{The Henry and Marilyn Taub Faculty of Computer Science, Technion, Israel}{amit.ganz@cs.technion.ac.il}{}{}
\author{Pranav Nuti}{Department of Mathematics, Stanford, USA}{pranavn@stanford.edu}{}{}
\author{Roy Schwartz}{The Henry and Marilyn Taub Faculty of Computer Science, Technion, Israel}{schwartz@cs.technion.ac.il}{}{}

\authorrunning{A. Ganz, P. Nuti and R. Schwartz} 

\Copyright{Amit Ganz, Pranav Nuti and Roy Schwartz} 



\funding{Amit Ganz and Roy Schwartz have received funding from the European Union’s Horizon 2020 research and innovation program under grant agreement no. 852870-ERC-SUBMODULAR}

\acknowledgements{The authors would like to thank the anonymous referees for helpful remarks.}

\begin{document}

\maketitle

\begin{abstract}
In this paper we consider the online Submodular Welfare (\SW) problem.
In this problem we are given $n$ bidders each equipped with a general non-negative (not necessarily monotone) submodular utility and $m$ items that arrive online. The goal is to assign each item, once it arrives, to a bidder or discard it, while maximizing the sum of utilities.
When an adversary determines the items' arrival order we present a simple randomized algorithm that achieves a tight competitive ratio of $\nicefrac{1}{4}$.
The algorithm is a specialization of an algorithm due to [Harshaw-Kazemi-Feldman-Karbasi MOR`22], who presented the previously best known competitive ratio of $3-2\sqrt{2}\approx 0.171573 $ to the problem.
When the items' arrival order is uniformly random, we present a competitive ratio of $\approx 0.27493$, improving the previously known $\nicefrac{1}{4}$ guarantee. Our approach for the latter result is based on a better analysis of the (offline) Residual Random Greedy (\RRG) algorithm of [Buchbinder-Feldman-Naor-Schwartz SODA`14], which we believe might be of independent interest.
\end{abstract}

\section{Introduction}\label{sec:intro}
Submodularity is a 
mathematical notion that captures the concept of diminishing returns. Formally, a set function $f:2^\mathcal{N} \rightarrow \mathbb{R}_{\geq 0}$ over a ground set $\mathcal{N}$ is submodular, if for all $A, B\subseteq \mathcal{N}$: $ f(A \cup B) + f(A \cap B) \leq f(A) + f(B) $.
An equivalent definition, which is called the diminishing returns property, is the following: $ f(A\cup \{ u\}) - f(A) \geq f(B\cup \{ u\})-f(B)$, for every $A\subseteq B\subseteq \mathcal{N}$ and every $u\in \mathcal{N}\setminus B$.
Submodular functions naturally arise in many different settings, {\em e.g.}, combinatorics, graph theory, information theory and economics.


We consider the Submodular Welfare (\SW) problem.
In this problem we are given a set $\mathcal{N}=\{ 1,\ldots,m\}$ of $m$ unsplittable items and a set $B=\{ 1,\ldots,n\}$ of $n$ bidders. Each bidder $j$ has a non-negative (and not necessarily monotone) submodular utility function $f_j$ and the goal is to assign items to the bidders while maximizing the sum of the utilities: $\sum_{j=1}^{n}f_j(S_j) $.
Here $S_j$ is the set of items allocated to bidder $j$, and the requirement is that $ S_j\cap S_{j'}=\emptyset$ for every $ j\neq j'$ (since the items are unsplittable) and $\cup _{j=1}^n S_j\subseteq \mathcal{N}$ (note that not all items must be assigned).
\SW with monotone utilities has been extensively studied for more than two decades, {\em e.g.},
\cite{CCPV07,CG10,DS06,FV06,KLMM05,LLN01,NS06,V08,V13}.
Submodular maximization with general (not necessarily monotone) objectives is also the focus of extensive theoretical research, {\em e.g.}, \cite{LMV10,LSV10,FMV11,FNS11b,VCZ11,GO11,V13,GP15,PJGBJ14,F17}.
Moreover, maximization of non-monotone submodular objectives has found numerous practical applications, {\em e.g.}, network inference \cite{HL17}, mobile crowdsensing \cite{LZWTC17}, summarization of documents and video \cite{LB10,LB11,MJK18}, marketing in social networks \cite{PNR15}, and even gang violence reduction \cite{SSPB14}, to name a few. 
In particular, non-monotone utilities in the context of \SW can model soft budget constraints, where each bidder $j$ needs to pay a price $ p_{i,j}$ when item $i$ is allocated to it.
Additional related problems were also studied, including: \SW with demand queries \cite{FV10}, and other utilities such as XOS and subadditive \cite{DNS10,Fei09}.


When considering the offline version, \SW is typically viewed as a special case of maximizing a submodular objective subject to a partition matroid independence constraint: $(1)$ the ground set is $\mathcal{N}\times B$; $(2)$ the partition matroid is defined over $\mathcal{N}\times B$ where each part in the partition corresponds to an item, {\em i.e.}, the part that corresponds to item $i\in \mathcal{N}$ is $\{ (i,j)\} _{j=1}^n$; and $(3)$ the objective $ f:2^{\mathcal{N}\times B}\rightarrow \mathbb{R}_{\geq 0}$
is defined as: $f(S)\triangleq \sum _{j=1}^n f_j(\{ i:(i,j)\in S \})$.
In case the utilities are monotone a tight (asymptotic) approximation of $(1-\nicefrac{1}{e})$ was given by \cite{CCPV11} by introducing the celebrated continuous greedy algorithm, a tight approximation of $(1-(1-\nicefrac{1}{n})^n)$ for any number $n$ of bidders was given by \cite{FNS11}, and the matching hardness result was given by \cite{BestSM}.
For a general and not necessarily monotone objective the current best known offline approximation is also based on the continuous approach and achieves an approximation of $(\nicefrac{1}{e})+0.0171$ \cite{BF19}, which improved upon the previous works of \cite{EN16,FNS11}.

In the online version of \SW items arrive one by one.
Whenever an item arrives, one has to decide immediately and irrevocably whether to assign it to one of the bidders or not assign it at all (the latter decision is relevant only when the utilities are not necessarily monotone).
There are two natural settings that differ in the order in which items arrive.
First, in the online adversarial setting an adversary can choose the order in which items arrive (the adversary knows the algorithm and how it operates, however if the algorithm is randomized it does not know the outcome of its random choices).
Second, in the online random order setting items arrive one by one in a uniform random order.

When considering the online version, as opposed to the offline version, worse results are known.
For monotone utilities, in the adversarial setting, a $(\nicefrac{1}{2})$-competitive greedy algorithm is known and additionally it is the best possible algorithm for this setting \cite{KPV12}.
However, if one assumes the online random order setting, it is known that one can achieve a competitive ratio of $(\nicefrac{1}{2})+0.0096$ \cite{BMG20}, which improved the result of \cite{GreedyBeats} who were the first to break the $\nicefrac{1}{2}$ barrier obtaining a competitive ratio of $(\nicefrac{1}{2})+0.005$. 

Unfortunately, when considering general submodular (and not necessarily monotone) utilities, worse results are known.
In the adversarial setting, the special case of a single bidder is of particular interest since it is equivalent to online Unconstrained Submodular Maximization (\USM): given a general submodular objective $f$ items arrive one by one in an online manner and once an item arrives the algorithm needs to decide whether to choose or discard it where the goal is to maximize $f(S)$ ($S$ denotes the chosen elements).
An (implicit) competitive ratio of $\nicefrac{1}{4}$ was given by \cite{FMV11} for online \USM (the algorithm simply chooses every element independently with a probability of half).
This result was proved to be tight \cite{BFS19}.
For the general case of multiple bidders an algorithm achieving a competitive ratio of $3-2\sqrt{2}\approx 0.171573$ was given by \cite{HCKFK22} (this algorithm handles a general matroid independence constraint and assumes elements of the ground set arrive in an online manner). 

In the random order setting,
an (implicit) result follows from analyzing the offline Residual Random Greedy (\RRG) algorithm of \cite{BFNS14} for maximizing a non-monotone submodular objective subject to a matroid independence constraint. 
When considering \SW, the \RRG algorithm operates as follows: it chooses a uniform random order over the items and goes over the items in this order, and assigns each item to the bidder with the highest marginal value (if this marginal value is negative then the item is discarded).
Thus, if the \RRG algorithm achieves an approximation of $\alpha$ in the offline setting, then the deterministic greedy achives a competitive ratio of $ \alpha$ in the online random order setting.
\cite{BFNS14} prove that the \RRG algorithm achieves an approximation of $\nicefrac{1}{4}$, therefore implying a competitive ratio of $\nicefrac{1}{4}$ for \SW in the random order setting. It is worth noting that better algorithms than \RRG are known in the offline setting, {\em e.g.}, \cite{BF19,EN16,FNS11}, however, they do not apply to the online random order setting.

In this work, we assume the standard {\em value oracle} model: each submodular function $f$ is not given explicitly but rather the algorithm can query for every subset $S$ the value of $f(S)$.
The running time of the algorithm is measured not only by the number of arithmetic operations, but also by the number of value queries.
In the online version, for both adversarial and random order settings, the algorithm can query subsets $S$ only of items that already arrived.
Thus, intuitively, the algorithm has no information regarding future items.

\subsection{Our Results}
Focusing first on the adversarial setting, we present the following positive result.
\begin{theorem}\label{thrm:SW_adv_alg}
 There exists a randomized polynomial time algorithm achieving a competitive ratio of $\nicefrac{1}{4}$ for online \SW in the adversarial setting with general (not necessarily monotone) utilities.
 \end{theorem}
There are three things to note regarding Theorem \ref{thrm:SW_adv_alg}.
First, the competitive ratio of $\nicefrac{1}{4}$ is tight as even for the special case of a single bidder (which is equivalent to online \USM) \cite{BFS19} provide a matching hardness of $\nicefrac{1}{4}$.
Second, Theorem \ref{thrm:SW_adv_alg}, to the best of our knowledge, improves the previous best known competitive ratio of $ 3-2\sqrt{2}\approx 0.171573$ \cite{HCKFK22}.
Third, for the special case of a single bidder the competitive ratio of Theorem \ref{thrm:SW_adv_alg} matches the $\nicefrac{1}{4}$ guarantee of the simple algorithm that just chooses independently for every item to include it in the solution with a probability of half \cite{FMV11}.
However, we note that the algorithm we present in order to prove Theorem \ref{thrm:SW_adv_alg}, even in the special case of a single bidder, differs from the algorithm that just chooses a uniform random subset.

We complement the above result by showing that the randomness of the online algorithm in Theorem \ref{thrm:SW_adv_alg} is needed.
This is summarized in the following theorem that gives a hardness result that tends to zero.
\begin{theorem}\label{thrm:SW_hard_det}
For every $ M>0$, no deterministic algorithm can achieve a competitive ratio better than $\nicefrac{1}{M}$ for online \SW in the adversarial setting with general (not necessarily monotone) utilities.
\end{theorem}

Focusing on the random order setting, the following theorem proves that one can achieve an improved competitive ratio over the previously (implicit) known $\nicefrac{1}{4}$ \cite{BFNS14}. 
We note that the following theorem separates the random order and adversarial settings, since one should recall there is a hardness of $\nicefrac{1}{4}$ in the adversarial setting even when only a single bidder is present.

\begin{theorem}\label{thrm:SW_alg}
The deterministic greedy algorithm
achieves a competitive ratio of $\approx 0.27493$ for online \SW in the random order setting with general (not necessarily monotone) utilities.
\end{theorem}



The above theorem is achieved by a better analysis of the offline randomized \RRG algorithm of  \cite{BFNS14}.
This improved analysis can be easily extended to a general matroid independence constraint, as the following theorem states (its proof appears in Section \ref{app:general_matroid}).


\begin{theorem}\label{thrm:RRG}
The \RRG algorithm of \cite{BFNS14} achieves an approximation guarantee of $\approx 0.27493$ for maximizing a general (not necessarily monotone) submodular function given a matroid independence constraint.
\end{theorem}

\subsection{Our Approach}\label{sec:Approach}

The online algorithm for the adversarial setting adopts a simple randomized approach: once item $i$ arrives it defines a distribution over the bidders and assigns the item to a random bidder sampled from this distribution.
Surprisingly, we prove that a remarkably simple distribution suffices to obtain a tight competitive ratio of $\nicefrac{1}{4}$:
item $i$ is assigned to the bidder with the highest marginal value with a probability of $\nicefrac{1}{2}$, to the bidder with the second highest marginal with a probability of $ \nicefrac{1}{4}$, and so forth as long as the marginal value is non-negative.
With the remainder probability item $i$ is discarded.
It is important to note that the {\em ordering} of the bidders according to their marginal values (as long as the marginal values are non-negative) dictates the distribution.
However, as long as the ordering remains the same, the distribution is independent of the actual marginal values themselves.
We prove that the above simple approach suffices to obtain a tight competitive ratio of $ \nicefrac{1}{4}$ for the adversarial setting.

It should be noted that the above randomized approach is based on the streaming algorithm of \cite{HCKFK22}, which obtains a competitive ratio of $ 3-2\sqrt{2}\approx0.171573$. 
Intuitively, we specialize the algorithm of \cite{HCKFK22} to online \SW in the adversarial setting.
The reason is that once item $i$ arrives one can perform the following process in order to obtain the distribution over bidders that was defined above: sort the bidders in a non-increasing order of marginal values and assign the item to the first bidder with a probability of half, if the item was not assigned to this bidder then with a probability of half assign the item to the next bidder, and so forth as long as the marginal value is non-negative.
This process describes how the algorithm of \cite{HCKFK22} operates in the special case the online order of elements of the ground set $ \mathcal{N}\times B$ satisfies: $(1)$ all elements $\{ (i,j)\}_{j=1}^n$ are consecutive in the online order; and $(2)$ elements $\{ (i,j)\}_{j=1}^n$ are sorted in a non-increasing order of marginal values of the bidders.

Focusing on the random order setting, we present an improved analysis of the \RRG algorithm for a general matroid independence constraint. 
As a preliminary step, which is not required but is mathematically convenient, we present a ``smooth'' version of the \RRG algorithm.
The difference between the original and smooth versions is that in the smooth version the distribution of the steps the algorithm can perform does not change as the algorithm progresses.
However, in the original version this distribution evolves as the algorithm progresses.
We note that the analysis of both versions is very similar (assuming the smooth version performs enough steps).
Nonetheless, we believe it is easier to analyze the smooth version.{\footnote{For a detailed comparison of the two versions see Section \ref{comparison}.}
%


A key insight in analyzing the \RRG algorithm (as well as other closely related algorithms, {\em e.g.}, the Random Greedy algorithm of \cite{BFNS14}), is lower bounding the expected value of a fixed optimal solution $OPT $ when a random subset of elements $S$ is added to it.
More precisely, if each element $u$ satisfies $\Pr [u\in S] \leq p$ then standard known arguments, {\em e.g.}, Lemma 2.2 \cite{BFNS14} which is based on Lemma 2.2 \cite{FMV11}, imply that $\mathbb{E}[f(OPT\cup S)] \geq (1-p)f(OPT)$.
It is important to note that this general insight works for {\em any} distribution of $S$ that satisfies $\Pr [u\in S] \leq p $ for every $u$. 
This general insight by itself enabled \cite{BFNS14} to prove that the \RRG algorithm achieves an approximation of $\nicefrac{1}{4}$.
We are able to improve upon the above by exploiting the specific probabilistic behavior of the smooth version of the algorithm, as opposed to the general insight which works for any distribution of $S$.
This results in an improved analysis of the smooth version of the \RRG algorithm via two jointly related recursive relations that together bound the performance of the algorithm.

\subsection{Additional Related Work}

The literature regarding submodular maximization, {\em i.e.}, problems of the form $ \max \{ f(S):S\in \mathcal{F}\}$ ($ \mathcal{F}$ is the collection of feasible solutions), is rich, {\em e.g.}, \cite{submodular,BestSM,CCPV11,KG05,S04,BV14,EN18,KMN99,KST13,BFNS15,FMV11,BFNS14,BF19,EN16}, and dates back to the late 70's.
Moreover, the online \SW problem naturally generalizes many online problems such as online matching \cite{KVV90,KP00,FMMM09,AGKM,MY11}, online weighted matching \cite{AGKM,FKMMP09}, budgeted allocation \cite{MSVV07,BJNHW07,DH09,GM08}, and more general classes of online allocation problems \cite{FHKMS10,VVS10,AWY14,DHKMY16}.
Other related problems include the extensively studied class of secretary problems, {\em e.g.}, \cite{Dyn63,Kle05,BIK07,Lac14,FSZ,BUCM12,BHZ13,FNS11secretary,GRST10,MTW16}, where the goal is to solve an optimization problem assuming the arrival order of the input is uniform and random.

\subsection{Preliminaries}\label{sec:prelim}


In our analysis we require the following known lemma.
\begin{lemma}[Lemma 2.2 \cite{BFNS14} which is based on Lemma 2.2 \cite{FMV11}] \label{lemma:p}
    Let $f: 2^\mathcal{N} \rightarrow \mathbb{R}_{\geq 0}$ be submodular. Denote by $A(p)$ a random subset of $A$ where each element appears with probability at most $p$ (not necessarily independently). Then $\mathbb{E}[f(A(p))] \geq (1-p)f(\emptyset)$.
\end{lemma}



\subsection{Paper Organization}
Section \ref{sec:adv} deals with the adversarial setting, proving Theorems \ref{thrm:SW_adv_alg} and \ref{thrm:SW_hard_det}.
Section \ref{sec:rand} focuses on the random order setting, proving Theorem \ref{thrm:SW_alg} and \ref{thrm:RRG}.

\section{Adversarial Setting}\label{sec:adv}
In this section we consider the adversarial setting, and present both a tight randomized algorithm proving Theorem \ref{thrm:SW_adv_alg} and hardness for any deterministic algorithm (Theorem \ref{thrm:SW_hard_det}).

\subsection{Tight Randomized Algorithm}\label{sec:adv_alg_rand}

In this section we consider the adversarial setting, and start by presenting our algorithm which appears in Algorithm \ref{alg:adv}.
For simplicity of presentation, we assume the items are numbered according to the order the adversary chooses, {\em i.e.}, item $1$ is the first to arrive, item $2$ is the second to arrive and so forth.
Hence, the algorithm performs $m$ iterations where in iteration $i$ the algorithm chooses what to do with item $i$: with a probability of $\nicefrac{1}{2}$ it assigns it to the bidder with highest marginal value, with a probability of $\nicefrac{1}{4}$ is assigns it to the bidder with the second highest marginal value, and so forth as long as the marginal value is non-negative.
With the remainder probability item $i$ is discarded and not assigned to any bidder.
Therefore, if there are $\ell$ bidders with non-negative marginal with respect to item $i$, item $i$ is discarded with a probability of $ 2^{-\ell}$.
In what follows we use the notation $f_j(i|S)$ to denote the marginal value of item $i$ with respect to bidder $j$ assuming bidder $j$ was already assigned a subset $ S\subseteq \mathcal{N}$ of items, {\em i.e.}, $ f_j(i|S)\triangleq f_j(S\cup \{i\})-f_j(S)$.



\RestyleAlgo{ruled}
\begin{algorithm}[hbt!]
\caption{}\label{alg:adv}
$\forall j=1,\ldots,n:$ $ S^0_j\gets  \emptyset$.\\
\For{$i=1,...,m$}{
$\forall j=1,\ldots ,n$: $ S^i_j\gets S^{i-1}_j$.\\
$\forall r=1,\ldots,n$: let $j_r$ be the bidder with the $r$\textsuperscript{th} highest marginal w.r.t item $i$:\\
~~~~~~~~~~~$ f_{j_1}(i|S^{i-1}_{j_1})\geq f_{j_2}(i|S^{i-1}_{j_2})\geq \ldots \geq f_{j_n}(i|S^{i-1}_{j_n})$.\\
let $j^i$ be a random bidder such that $ \Pr[j^i=j_r]=2^{-r}$.\\
\If {$f_{j^i}(i|S^{i-1}_{j^i}) \geq 0$} {
$S^i_{j^i}\leftarrow S^i_{j^i}\cup \{i\}$\\}}
return $S^m_1, S^m_2 , \dots, S^m_n$.\\
\end{algorithm}


Let $O$ denote an offline optimal allocation for the problem: $O = (O_1 , \dots, O_n)$, where $O_j$ is the collection of items assigned to bidder $j$ in the optimal solution.
Formally, $O$ is defined as follows\footnote{Note that formally speaking this is actually a set, but for notational convenience, in this paper, when we refer to something as equalling the $\argmax$, we will always mean it equals an element of the $\argmax$ set.}:
$$\argmax _{(O_1,\ldots,O_n)}\left\{\sum_{j=1}^{n}f_j(O_j):\forall j\neq r ~~~O_j\cap O_r=\emptyset,~~~\forall j=1,\ldots,n ~~~O_j\subseteq \mathcal{N}\right\}.$$

Let $S^i$ be the allocation induced by the algorithm at the end of the $i$\textsuperscript{th} iteration, {\em i.e.}, $ S^i=( S^i_1,S^i_2,\ldots,S^i_n)$. For every bidder $j$, define $H^i_j$ as follows:
$$ H^i_j \triangleq (O_j \cap \{1, \ldots, i\}) \cup S^i_j,$$
and let $H^i=(H^i_1,\ldots,H^i_n)$.
It is important to note that $H^i$ is not necessarily a feasible solution, since an item might be assigned to up to two bidders.
For simplicity of presentation we use the notation of $f(S^i)$ to denote the value of the allocation $S^i$, {\em i.e.}, $ f(S^i)\triangleq \sum _{j=1}^n f_j(S^i_j)$.
Similarly, we use $f(H^i)$ to denote $ \sum _{j=1}^n f_j(H^i_j)$ (though $H^i$ is not an allocation since items might be assigned to more than a single bidder).

In order to analyze the algorithm we denote by $P^i$ the profit gained during the $i$\textsuperscript{th} iteration: $P^i \triangleq f(S^i)-f(S^{i-1})$.
Note that $P^1+\ldots + P^m =f(S^m)-f(S^0)$, where $S^m$ is the output of Algorithm \ref{alg:adv} and $S^0$ is the empty allocation that does not assign any item to any of the bidders.
Finally, we define the sequence: $K^i \triangleq f(H^{i})-f(H^{i-1})$.
Note that $K^1+\ldots+K^m=f(H^m)-f(S^0)$ (note that $ H^0=S^0$).

The analysis of the competitive ratio of Algorithm \ref{alg:adv} is essentially based on a single observation, which for every iteration upper bounds $K^i$ as a function of the expected gained profit.
This is summarized in Lemma \ref{lem:adversary}.

\begin{lemma}\label{lem:adversary}
$\forall i=1,\ldots,m$ the following holds:
$ \mathbb{E}[K^i] \leq 2\cdot \mathbb{E}[ P^i]$.
\end{lemma}

We use Lemma \ref{lem:adversary} to prove Theorem \ref{thrm:SW_adv_alg}. The proof is essentially by summation over all iterations of the algorithm.

\begin{proof}[Proof of Theorem \ref{thrm:SW_adv_alg}]We prove that the competitive ratio of Algorithm \ref{alg:adv} is $\nicefrac{1}{4}$. 
    Lemma \ref{lem:adversary}, when applied for every iteration $i$, implies that: $\mathbb{E}\left[\sum_{i=1}^m K^i\right] \leq 2 \cdot \mathbb{E}\left[\sum_{i=1}^m P^i\right]$.
    Hence, it follows that: 
    $\mathbb{E}\left[f(H^m)-f(S^0)\right] \leq 2 \cdot \mathbb{E}\left[f(S^m)-f(S^0)\right]$.
    In fact, since $f(S^0)$ is non-negative, we have that: 
    $$\mathbb{E}\left[f(H^m)\right] \leq 2 \cdot \mathbb{E}\left[f(S^m)\right] .$$
    For every bidder $j$ and item $i$ we have that the probability that item $i$ is assigned to bidder $j$ is at most $\nicefrac{1}{2}$, {\em i.e.}, $\Pr[i\in S_j^m] \leq \nicefrac{1}{2}$.
    Therefore, using Lemma \ref{lemma:p} applied to the submodular function $h_j$, where $ h_j:2^{\mathcal{N}}\rightarrow \mathbb{R}_{\geq 0}$ and $ h_j(S)\triangleq f_j(S\cup O_j)$ for every $S\subseteq \mathcal{N} $, we can conclude that: $\mathbb{E}[f_j(S^m_j\cup O_j)]=\mathbb{E}[h_j(S^m_j)] \geq (\nicefrac{1}{2})\cdot h_j(\emptyset)=(\nicefrac{1}{2})\cdot f_j(O_j) $.
    Therefore,
    $$\mathbb{E}[f(H^m)]=\sum _{j=1}^n \mathbb{E}[f_j(H^m_j)]=\sum _{j=1}^n\mathbb{E}[f_j(S_j^m\cup O_j)]
    \geq \frac{1}{2}\sum _{j=1}^n f_j(O_j)=\frac{1}{2}\cdot f(O).$$
    Combining the above, we conclude that: $(\nicefrac{1}{4}) \cdot f(O) \leq \mathbb{E}[f(S^m)] 
    $.
\end{proof}

All that remains is to prove Lemma \ref{lem:adversary}.

\begin{proof}[Proof of Lemma \ref{lem:adversary}]
For proof simplicity, let us start by adding a dummy bidder whose utility is the zero function. Clearly, this does not change the algorithm's performance or the value of $f(O)$, but it lets us assume (without loss of generality) that: $(1)$ for every item $i$ the optimal solution $O$ allocates item $i$ to some bidder $k$, {\em i.e.}, $i\in O_k$; and $(2)$ for every item $i$ at least one bidder has a non-negative marginal value, {\em i.e.}, $ f_{j}(i|S^{i-1}_{j})\geq 0$ for some bidder $j$. 

Fix an iteration $i=1,\ldots, m$ and condition on any possible realization $R_{i-1}$ of the random choices of the algorithm in the first $i-1$ iterations. Thus, $S^{i-1}$, $H^{i-1}$,  and $ j_1$ up to $j_m$ are deterministic and fixed given this conditioning, whereas $ j^i$ is a random variable.

Recall that (without loss of generality) item $i$ is assigned to bidder $k$ by the optimal solution and that $j_r$ denotes the bidder with the $r^\text{th}$ highest marginal value with respect to item $i$ given the elements previously assigned to the bidder. Let us assume that the first $\ell\geq 1$ bidders have a non-negative marginal value with respect to item $i$, {\em i.e.}, $ f_{j_1}(i|S^{i-1}_{j_1})\geq f_{j_2}(i|S^{i-1}_{j_2}) \geq \ldots \geq f_{j_{\ell}}(i|S^{i-1}_{j_{\ell}})\geq 0$ and if $ \ell < m$: $ f_{j_{\ell +1}}(i|S^{i-1}_{j_{\ell +1}})<0$. 
Moreover, assume that bidder $k$ has the $t$\textsuperscript{th} largest marginal, {\em i.e.}, $ k=j_t$.
Let us now bound the expected change from $f(H^{i-1})$ to $\mathbb{E}[f(H^i)|R_{i-1}] $, {\em i.e.}, $ \mathbb{E}[K^i|R_{i-1}]$, given the assumption that $ t\leq \ell$, {\em i.e.}, bidder $k$ to which the optimal solution assigned item $i$ is among the $\ell$ bidders who have a non-negative marginal value: 
\begin{align}
 & \mathbb{E}[f(H^i)|R_{i-1}] - f(H^{i-1})\nonumber\\
 = &f_k(S_k^{i-1}\cup (O_k\cap \{ 1,\ldots, i-1\})\cup \{ i\}) - f_k(S_k^{i-1}\cup (O_k\cap \{ 1,\ldots,i-1\})) + \label{adv_1}  \\
 &{\sum_{r = 1}^{\ell}\frac{\mathbf{1}_{\{j_r \neq k\}}}{2^{r}} \left\{ f_{j_r}((O_{j_r}\cap \{ 1,\ldots,i-1\})\cup S_{j_r}^{i-1}\cup \{i\}) - f_{j_r}((O_{j_r}\cap \{ 1,\ldots,i-1\})\cup S_{j_r}^{i-1})\right\}} .\nonumber
\end{align}
The equality in (\ref{adv_1}) follows from the definitions of $H^i$ and the algorithm, as well as the fact that $ i\in O_k$.
We note that that (\ref{adv_1}) can be upper bounded as follows:
\begin{align}
 \leq & f_k(i|S_k^{i-1}) + {\sum_{r = 1}^{\ell}\frac{\mathbf{1}_{\{j_r \neq k\}}f_{j_r}(i|S_{j_r}^{i-1})}{2^r}}. \label{adv_2}
\end{align}
The inequality in (\ref{adv_2}) follows from the decreasing marginals property of the submodular utilities.
Next, let us rewrite (\ref{adv_2}): 
\begin{align}
 = & f_{j_t}(i|S_{j_t}^{i-1})\cdot\left(1-\frac{1}{2^t}\right)+ {\sum_{r = 1}^{\ell}\frac{f_{j_r}(i|S_{j_r}^{i-1})}{2^r}}\label{adv_3}\\
 = & f_{j_t}(i|S_{j_t}^{i-1}) \cdot\left(\sum _{r=1}^t 2^{-r} \right)+ {\sum_{r = 1}^{\ell}\frac{f_{j_r}(i|S_{j_r}^{i-1})}{2^r}}.\label{adv_4}
 \end{align}
 The equality in (\ref{adv_3}) holds since bidder $k$ has the $t$\textsuperscript{th} largest marginal, {\em i.e.}, $ k=j_t$, and $ t\leq \ell$.
Also, the equality in (\ref{adv_4}) follows from the value of a geometric sum.
Moreover, we note that (\ref{adv_4}) can be further upper bounded:
\begin{align}
 \leq & {\sum_{r = 1}^{t}\frac{f_{j_r}(i|S_{j_r}^{i-1})}{2^r}} + {\sum_{r = 1}^{\ell}\frac{f_{j_r}(i|S_{j_r}^{i-1})}{2^r}}. \label{adv_5}
 \end{align}
In the above, the inequality in (\ref{adv_5}) is true since bidder $j_r$ has the $r$\textsuperscript{th} largest marginal value, {\em i.e.}, for every $ r\leq t$: $ f_{j_r}(i|S^{i-1}_{j_r})) \geq f_{j_t}(i|S^{i-1}_{j_t})$.
Next, we upper bound (\ref{adv_5}) as follows:
\begin{align}
 \leq & 2 \cdot {\sum_{r = 1}^{\ell}\frac{f_{j_r}(i|S_{j_r}^{i-1})}{2^r}} .\label{adv_6}
\end{align}
We note that the inequality in (\ref{adv_6}) follows since we assumed $ t\leq \ell$, {\em i.e.}, bidder $k$ is among the $\ell$ bidders with non-negative marginal values.
Hence, the first sum in (\ref{adv_5}) can be extended to include all $ \ell$ bidders with non-negative marginal value.
Finally, we note that (\ref{adv_6}) equals the following by the definition of Algorithm \ref{alg:adv}:
\begin{align}
 = & 2\cdot \mathbb{E}[P^i|R_{i-1}]. \nonumber
\end{align}
Hence, we can conclude that $ \mathbb{E}[K^i|R_{i-1}]\leq 2\cdot \mathbb{E}[P^i|R_{i-1}]$ as desired.

We note that if $ t>\ell$ then the latter inequality trivially holds (the above proof works until (\ref{adv_2}) in which the first term is negative and thus can be dropped which implies that $ \mathbb{E}[K^i|R_{i-1}]\leq \mathbb{E}[P^i|R_{i-1}]$  and hence that $ \mathbb{E}[K^i|R_{i-1}]\leq 2\cdot \mathbb{E}[P^i|R_{i-1}]$, since $P_i$ is non-negative).
Thus, using the law of total expectation over all possible outcomes $ R_{i-1}$ the proof is complete.
\end{proof}

\subsection{Deterministic Hardness}\label{sec:adv_hard}
\begin{proof}[Proof of Theorem \ref{thrm:SW_hard_det}]

We present an instance for which any online deterministic algorithm cannot achieve a competitive ratio better than $\nicefrac{1}{M}$ for every $M>0$ versus an adversary.
We consider an instance with a single bidder $B=\{b\}$ and two items $\mathcal{N} = \{v_1,v_2\}$.
The items arrive according to their index in an online manner, {\em i.e.}, item $v_1$ is the first to arrive, and item $v_2$ is the second to arrive.
Once item $v_1$ arrives, any deterministic algorithm can query  $f $ on subsets of elements that can contain only $v_1$.
Hence, we only need to define $f$ on $\emptyset$ and $\{ v_1\}$.
We define $f$ as follows: $f(\emptyset) = 0, f(\{v_1\}) = 1$.
How this utility function is extended to a submodular function over all subsets of items depends on what the algorithm chooses to do once item $v_1$ arrives.

The first case is when the algorithm chooses not to assign $v_1$ to $b$. 
In this case we extend the above definition of $f$ by setting the contribution of $v_2$ to be linearly zero, {\em i.e,}, $ f(\{v_2\})\triangleq 0$ and $f(\{v_1, v_2\})\triangleq 1$.
One can verify that the resulting utility function $f$ is indeed submodular.
In this case, the value of an optimal solution to the instance equals $1$, since item $v_1$ can be assigned to bidder $b$ by the optimal solution.
However, every deterministic algorithm that does not assign $v_1$ to $b$ has a value of $0$.
Thus, we conclude a competitive ratio of $0$ in this case.

The second case is when the algorithm chooses to assign $v_1$ to bidder $b$.
In this case we extend the above definition of $f$ as follows: $f(\{v_2\})\triangleq M$ and $f(\{v_1, v_2\})\triangleq 0$.
One can verify that the resulting utility function $f$ is indeed submodular.
In this case, the value of an optimal solution to the instance equals $M$, since the optimal solution can choose to assign only $\{v_2\}$ to $b$.
However, every deterministic algorithm that assigns $v_1$ to bidder $b$ achieves a value of at most $1$.
Thus, we conclude a competitive ratio of at most $\nicefrac{1}{M}$ in this case.

In conclusion, for this instance, no deterministic algorithm can achieve any competitive ratio better than $\nicefrac{1}{M}$ in the online adversarial setting, for every constant $M$.
\end{proof}

\section{Uniform Random Order Setting}\label{sec:rand}
In this section we focus on the uniform random order setting. 
Recall that it is known that if the \RRG algorithm provides an approximation of $\alpha$ for maximizing a general submodular function given a partition matroid independence constraint, then it also provides an online algorithm in the random order setting which achieves a competitive ratio of $\alpha$ (see brief discussion in Section \ref{sec:intro}).

To simplify the presentation of our improved analysis, we present a ``smooth'' version of the \RRG algorithm of \cite{BFNS14}.
Though the analysis of the smoothed version is similar to the original \RRG (assuming enough iterations are performed), we believe it is easier to analyze since all iterations have the same probabilistic distribution (whereas in the original \RRG this is not the case).
For simplicity of presentation, we first focus on the case the matroid is a partition matroid.
Recall that this special case already captures \SW.

\subsection{Partition Matroid}
We are given a partition matroid $\mathcal{M}=(\mathcal{N},\mathcal{I})$ over a ground set $\mathcal{N}$ which is partitioned into disjoint non-empty sets $P_1, P_2, \ldots ,P_k$. The goal is to choose a subset $S \subseteq \mathcal{I}$, {\em i.e.}, $S$ contains at most one element from each set $P_j$, that maximizes a given non-negative (general) submodular function $f$. 

We call our algorithm smooth since the random choices of the algorithm are always uniform, no matter how many iterations the algorithm performed so far.
Specifically, the smooth algorithm chooses in every iteration a part uniformly at random from {\em all} $k$ parts of $\mathcal{N}$, and adds the best element in the chosen part assuming that part does not intersect what the algorithm chose so far.
The original \RRG chooses a part uniformly at random from parts that do not intersect what the algorithm chose so far, and adds the best element in the chosen part.
Thus, the number of iterations the smooth algorithm can perform is unlimited.
A formal description of the algorithm for partition matroids is given as Algorithm \ref{alg:partition}, and we note that the number of iterations $T$ is a parameter given to the algorithm.
For a comparison of Algorithm \ref{alg:partition} and the original \RRG refer to section \ref{comparison}.

Given any $S \subseteq \mathcal{N}$, we denote the parts of the partition of  $\mathcal{N}$ that do not intersect $S$ by $I(S)$ ,{\em i.e.}, $I(S)\triangleq \left\{j=1,...,k | P_j \cap S = \emptyset \right\}$.
Without loss of generality, we assume every part $P_j$ is padded with a dummy element (a different dummy element for every $P_j$) that linearly contributes zero to the objective $f$. Hence, without loss of generality, $|OPT| = k$, where we denote by $OPT$ an optimal solution to the problem: $OPT \triangleq \argmax\{f(S)|S\in \mathcal{I}\}$.

\RestyleAlgo{ruled}
\begin{algorithm}[hbt!]
\caption{Smooth Residual Random Greedy (Partition Matroid)}\label{alg:partition}
$S_0 \gets \emptyset$.\\
\For{$i=1,...,T$}{
    $S_i \leftarrow S_{i-1}$.\\
    $M_i  \gets \bigcup_{j\in I(S_{i-1})}\left\{\argmax\left\{f\left(S_{i-1}\cup \{u\}\right)-f(S_{i-1})\right | u\in P_j\}\right\}$.\\
    Let $j$ be a uniformly random number from $\{1,...,k\}$.\\
    \If{$j\in I(S_{i-1})$}{
    Let $u_i$ be the element from $P_j$ in $M_i$ and $ S_i\leftarrow S_i\cup \{ u_i\}$.\\
    }
}
  Return $S_T$.
\end{algorithm}

For every set $S \in \mathcal{I}$ we denote $O_{S} \triangleq  \argmax_A\{f(S\cup A)|S\cup A\in \mathcal{I}, A\subseteq \mathcal{N}\setminus S\}$, {\em i.e.}, $O_{S}$ is the best extension of $S$ to an independent set.
Recalling that every part $P_j$ is padded with a dummy element that linearly contributes zero to the objective implies that $|O_S|=k-|S|$. 
One should note that $ O_{\emptyset}=OPT$, and thus $O_{S_0}=OPT$.
Our analysis tracks $f(O_{S_i}\cup S_i)$ as $S_i$ changes throughout the algorithm.
Intuitively, $f(O_{S_i}\cup S_i)$ deteriorate as more elements are added to $S_i$.
Building on the above intuition, the following two lemmas establish a system of joint recursive formulas for $\mathbb{E}[f(S_i)]$ and $\mathbb{E}\left[f\left(O_{S_i}\cup S_{i}\right)\right]$.

\begin{lemma}\label{lemma:exp of s}
For every $i=1,\ldots,T$:
$$\mathbb{E}\left[f(S_i)\right] - \mathbb{E}\left[f(S_{i-1})\right]\geq \frac{1}{k}\cdot \mathbb{E}\left[f\left(O_{S_{i-1}}\cup S_{i-1}\right)-f(S_{i-1})\right].$$
\end{lemma}

\begin{proof}
Fix $i=1,\ldots,T$ and condition on any possible realization of the choices of the algorithm in the first $i-1$ iterations. Thus, $S_{i-1}$, $M_i$, and $O_{S_{i-1}}$ are deterministic and fixed given this conditioning, and $u_i$ and $S_i$ are the only random variables. For the remainder of the proof all the probabilities and expectations are conditioned on this possible realization.

\begin{align}
\mathbb{E}\left[f(S_i)\right] - f(S_{i-1}) & = \frac{1}{k} \sum_{u\in M_i} \left\{f(S_{i-1} \cup \{u\})-f(S_{i-1})\right\}\label{ineq new} \\
 & \geq \frac{1}{k} \sum_{u\in O_{S_{i-1}}} \left\{f\left(S_{i-1} \cup \{u\}\right)-f(S_{i-1})\right\}\label{ineq1} \\
 & \geq \frac{1}{k} \left\{f(S_{i-1} \cup O_{S_{i-1}})-f(S_{i-1})\right\} \label{ineq2}
\end{align}

In the above, the equality in (\ref{ineq new}) follows from the algorithm's definition.
The inequality in (\ref{ineq1}) follows from the greedy choice of $M_i$, and the inequality in (\ref{ineq2}) follows from the submodularity of $f$.
We conclude the proof by unfixing the conditioning and taking an expectation over all possible such events (the law of total expectation).
\end{proof}

The following lemma provides a recursive formula for $ \mathbb{E}[f(O_{S_i}\cup S_i)]$ whose novelty is in the added contribution of $f(S_{i-1})$.
Without the added $\mathbb{E}\left[ f(S_{i-1})\right] /k$ term the resulting approximation will be $\nicefrac{1}{4}$ as in \cite{BFNS14}.
\begin{lemma}\label{lemma:exp of xor}
For every $i=1,\ldots,T$:
$$\mathbb{E}\left[f\left(O_{ S_i}\cup S_i\right)\right] \geq \left( 1-\frac{2}{k}\right) \cdot \mathbb{E}\left[f\left(O_{S_{i-1}}\cup S_{i-1}\right)\right]+\frac{1}{k}\cdot\mathbb{E}\left[ f(S_{i-1})\right].$$ 
\end{lemma}
\begin{proof}
Fix $i=1,\ldots,T$ and condition on any possible realization of the choices of the algorithm in the first $i-1$ iterations. Thus, $S_{i-1}$, $M_i$, and $O_{S_{i-1}}$ are deterministic and fixed given this conditioning, and $u_i$, $S_i$, and $ O_{S_i}$ are the only random variables. For the remainder of the proof all the probabilities and expectations are conditioned on this possible realization. 

For every element $u \in M_i$ we denote by $P(u)$ the index of the part of the partition of $\mathcal{N}$ which $u$ belongs to. Formally, $P(u) = j$ if and only if $u\in P_j$. We define $h:M_i \xrightarrow{} O_{S_{i-1}} $ to be a bijection mapping every element $u \in M_i$ to an element of $O_{S_{i-1}} $ in such a way that $P(u)= P(h(u))$.
One should note that our assumption that every $P_j$ contains a dummy element that contributes zero to the objective implies that $ I(S_{i-1})=I(O_{S_{i-1}})$, and hence $h$ is well defined.
Thus,
\begin{align}
&\mathbb{E}\left[f\left(O_{ S_i}\cup S_{i}\right)\right] - f\left(O_{S_{i-1}}\cup S_{i-1}\right) \nonumber\\
=&\frac{1}{k} \sum_{u\in M_i} \{f\left(S_{i-1} \cup \{u\} \cup O_{S_{i-1}\cup \{u\}}\right)-f \left(O_{S_{i-1}}\cup S_{i-1}\right)\} \label{eq:1} \\
\geq & \frac{1}{k} \sum_{u\in M_i} \{f\left(S_{i-1} \cup \{u\} \cup ( O_{S_{i-1}} \setminus \{h(u)\} )\right)-f \left(O_{S_{i-1}}\cup S_{i-1}\right)\} . \label{Os1}
\end{align}
In the above, the equality in (\ref{eq:1}) follows from the algorithm's definition.
The inequality in (\ref{Os1}) follows from the observation that $O_{S_{i-1}\cup \{u\}}$ is the best extension of $S_{i-1}\cup \{ u\}$ whereas $O_{S_{i-1}}\setminus \{ h(u)\}$ is just an extension of $S_{i-1}\cup \{u\}$, {\em i.e.},
$$ f(S_{i-1}\cup \{u\}\cup O_{S_{i-1}\cup \{u\}}) \geq f(S_{i-1}\cup \{u\}\cup (O_{S_{i-1}}\setminus \{ h(u)\})).$$
We note that (\ref{Os1}) equals the following:
\begin{align}
= &\frac{1}{k} \sum_{u\in M_i | u \neq h(u)} \{f\left(S_{i-1} \cup \{u\} \cup ( O_{S_{i-1}} \setminus \{h(u)\} )\right)-f \left(O_{S_{i-1}}\cup S_{i-1}\right)\} . \label{Os2}
\end{align}
The equality in (\ref{Os2}) holds since if $ u=h(u)$ then $ \{u\}\cup (O_{S_{i-1}}\setminus \{h(u)\}) = O_{S_{i-1}}$, and therefore summation can be reduced to all candidate elements $ u\in M_i$ satisfying $u\neq h(u) $.
Moreover, we note that (\ref{Os2}) can be lower bounded as follows:
\begin{align}
\geq &\frac{1}{k} \sum_{u\in M_i | u \neq h(u)} \{f\left(S_{i-1} \cup \{u\} \cup  O_{S_{i-1}} \right)-f\left(S_{i-1} \cup O_{S_{i-1}}\right)\} + \label{Os3} \\ 
& \frac{1}{k} \sum_{u\in M_i | u \neq h(u)} \{f\left(S_{i-1} \cup ( O_{S_{i-1}} \setminus \{h(u)\}) \right)-f\left(S_{i-1} \cup O_{S_{i-1}}\right)\} .\nonumber
\end{align}
The inequality in (\ref{Os3}) follows from submodularity since summation is restricted only to candidates $ u\in M_i$ satisfying $ u\neq h(u)$ and hence:
$ f(S_{i-1}\cup \{u\}\cup (O_{S_{i-1}}\setminus\{ h(u)\}))+f(S_{i-1}\cup O_{S_i-1}) \geq f(S_{i-1}\cup \{ u\}\cup O_{S_{i-1}}) +f(S_{i-1}\cup (O_{S_{i-1}}\setminus\{h(u)\}))$.

We further lower bound (\ref{Os3}) in the following way:
\begin{align}
\geq & \frac{1}{k} \sum_{u\in M_i } \{f\left(S_{i-1} \cup \{u\} \cup  O_{S_{i-1}} \right)-f\left(S_{i-1} \cup O_{S_{i-1}}\right)\} + \label{Os3.5} \\ 
& \frac{1}{k} \sum_{u\in M_i } \{f\left(S_{i-1} \cup ( O_{S_{i-1}} \setminus \{h(u)\}) \right)-f\left(S_{i-1} \cup O_{S_{i-1}}\right)\} .\nonumber
\end{align}
When examining the inequality in (\ref{Os3.5}) let us start with the first sum.
We note that if $ u=h(u)$ for some $ u\in M_i$ then $ u\in O_{S_{i-1}}$, {\em i.e.}, $ \{ u\}\cup O_{S_{i-1}}=O_{S_{i-1}}$.
Thus, extending the first sum to all $ u\in M_i$ (regardless of whether $ u$ equals $h(u)$ or not) does not change the first sum.
Focusing on the second sum, we note that for every $ u\in M_i$, regardless of whether $u$ equals $h(u)$ or not, the following holds: $ f(S_{i-1}\cup (O_{S_{i-1}}\setminus \{ h(u)\}))\leq f(S_{i-1}\cup O_{S_{i-1}})$.
The reason for the latter is that $ O_{S_{i-1}}$ is the best extension of $S_{i-1}$ whereas $ O_{S_{i-1}}\setminus \{h(u)\}$ is some extension of $S_{i-1}$.
Hence, adding to the second sum all terms corresponding to $u\in M_i$, where $ u=h(u)$, can only decrease the second sum.
Therefore, we conclude that the inequality in (\ref{Os3.5}) holds.
Finally, we lower bound (\ref{Os3.5}) as follows:
\begin{align}
\geq & \frac{1}{k} \{f\left(S_{i-1} \cup O_{S_{i-1}} \cup M_i\right)-f\left(S_{i-1} \cup O_{S_{i-1}}\right)\} + \label{Os5} \\
& \frac{1}{k} \{f\left(S_{i-1} \cup O_{S_{i-1}} \setminus O_{S_{i-1}}\right)-f\left(S_{i-1} \cup O_{S_{i-1}}\right)\} 
\nonumber \\
\geq &  \frac{1}{k} \{f(S_{i-1}) -2 \cdot \left(f\left(S_{i-1} \cup O_{S_{i-1}}\right)\right) \} \label{Os6}
\end{align}

The inequality in (\ref{Os5}) follows from submodularity which implies the following two: $$ \sum_{u\in M_i} \{f\left(S_{i-1} \cup \{u\} \cup O_{S_{i-1}} \right)-f\left(S_{i-1} \cup O_{S_{i-1}}\right)\} \geq f\left(S_{i-1} \cup O_{S_{i-1}} \cup M_i\right)-f\left(S_{i-1} \cup O_{S_{i-1}}\right)$$
$$ \sum_{u\in M_i} \{f\left(S_{i-1} \cup O_{S_{i-1}} \setminus \{h(u)\} \right)-f\left(S_{i-1} \cup O_{S_{i-1}}\right)\} \geq f\left(S_{i-1} \cup O_{S_{i-1}} \setminus O_{S_{i-1}}\right)-f\left(S_{i-1} \cup O_{S_{i-1}}\right) .$$
We note that the inequality in (\ref{Os6}) follows from the non-negativity of $f$ which implies that: $f\left(S_{i-1} \cup O_{S_{i-1}} \cup M_i\right) \geq 0$ and the fact that $ S_{i-1}\cup O_{ S_{i-1}} \setminus O_{ S_{i-1}}=S_{i-1}$.

We conclude the proof by unfixing the conditioning and taking an expectation over all possible such events (the law of total expectation). 

\end{proof}

The following lemma lower bounds the solution to the system of joint recursive formulas presented in Lemmas \ref{lemma:exp of s} and \ref{lemma:exp of xor} (its proof appears in Appendix \ref{app:lem_both}).

For simplicity of presentation we introduce two absolute constants: $ a\triangleq (3-\sqrt{5})/2\approx 0.381966$ and $ b\triangleq (3+\sqrt{5})/2\approx 2.61803$.
\begin{lemma} \label{lemma:both}
For every $i=0,1,\ldots,T$ the following hold:
 \begin{align}
 \mathbb{E}[f(S_i)] & \geq  \frac{f(OPT)}{\sqrt{5}}\left( \left( 1-\frac{a}{k}\right) ^i- \left( 1-\frac{b}{k} \right)^i\right)\label{ineq::1}\\
 \mathbb{E}\left[f(O_{S_i}\cup S_i )\right] & \geq \frac{f(OPT)}{2\sqrt{5}}\left( (\sqrt{5}-1)\left(1-\frac{a}{k} \right)^i+(\sqrt{5}+1)\left(1-\frac{b}{k} \right)^i\right). \label{ineq::2}
 \end{align} 
\end{lemma}


The following lemma establishes the approximation guarantee of Algorithm \ref{alg:partition}.
\begin{lemma}\label{lemme:alg:partition}
Algorithm \ref{alg:partition} achieves an approximation ratio of at least $0.27493$ for the problem of maximizing a non-monotone submodular function subject to a partition matroid independence constraint.
\end{lemma}
\begin{proof}
By our assumption of the existence of dummy elements, no element of $M_i$ has a negative marginal value, in any iteration $i$. We get that always for every $i$, $f(S_i) \geq f(S_{i-1})$ (note that this inequality holds for the random variables $S_i$ and $S_{i-1}$). Therefore, it suffices to show that $\mathbb{E}[f(S_i)] \geq 0.27493 \cdot f(OPT)$ for some $i=1,\ldots,T$.

Observe that setting $i=x^*k$, where $x^*=\ln(b/a)/\sqrt{5}=\ln{\left((3+\sqrt{5})/(3-\sqrt{5})\right)}/\sqrt{5}  \approx 0.86$, alongside Lemma \ref{lemma:both}, implies that:
\begin{align}
    \mathbb{E}[f(S_{x^*k})] \geq  \frac{f(OPT)}{\sqrt{5}}\left( \left( 1-\frac{a}{k}\right)^{x^*k}-\left( 1-\frac{b}{k}\right)^{x^*k}\right)\geq  \frac{f(OPT)}{\sqrt{5}}\left( e^{-x^*a}-e^{-x^*b}\right).\label{approx_ineq1}
\end{align}
The inequality above follows from the observation that $\left( 1-a/k\right) ^{x^*k} - \left( 1-b/k\right) ^{x^*k}$, as a function of $k$, is monotone decreasing for every $ k\geq 3$ (the proof of this technical observation is omitted).
Thus, the inequality follows by taking the limit $ k\rightarrow \infty$.
The lemma follows by plugging in the above values of $a$, $b$, and $x^*$.
\end{proof}



\begin{proof}[Proof of Theorem \ref{thrm:SW_alg}]
Follows immediately from Lemma \ref{lemme:alg:partition}.
\end{proof}

\subsection{General Matroid} \label{app:general_matroid}
In order to extend Algorithm \ref{alg:partition} to a general matroid we require the definition of a contracted matroid.
Given a matroid $\mathcal{M}=(\mathcal{N},\mathcal{I})$ and a set $S \in
\mathcal{I}$, the contracted matroid $\mathcal{M}/S \triangleq (\mathcal{N}\setminus S, \mathcal{I}_{\mathcal{M}/S})$ is defined as follows: a set $S'\subseteq \mathcal{N}\setminus S$ belongs to $\mathcal{I}_{\mathcal{M}/S}$ if and only if $S'\cup S \in \mathcal{I}$.
A formal description of the algorithm for a general matroid is given by Algorithm \ref{alg:matroid}.
It is important to note that Algorithm \ref{alg:matroid} reduces to Algorithm \ref{alg:partition} in the case that $\mathcal{M}$ is a partition matroid.
Similar to the case of a partition matroid, we can assume without loss of generality that $\mathcal{N}$ is padded with sufficient dummy elements that linearly contribute zero to the objective $f$ such that $|OPT| = k$, {\em i.e.}, $OPT$ is a base of $\mathcal{M}$.\footnote{One method to pad the ground set $\mathcal{N}$ is as follows: the new ground set is $ \mathcal{N}\cup \mathcal{N}'$ (where $ \mathcal{N}'=\{ u':u\in \mathcal{N}\}$), the matroid is $ \mathcal{M}'=(\mathcal{N}\cup \mathcal{N}',\mathcal{I}')$ where $ \mathcal{I}'=\{ S\subseteq \mathcal{N}\cup\mathcal{N}':\forall u\in \mathcal{N}~ |\{ u,u'\}\cap S|\leq 1, \{ u\in \mathcal{N}:u\in S \text{ or }u'\in S\}\in \mathcal{I}\}$, and all elements in $\mathcal{N}'$ contribute linearly zero to the submodular function, {\em i.e.}, $f'(S)=f(S\cap \mathcal{N})$ for every $ S\in \mathcal{N}\cup \mathcal{N}'$. 
}

\RestyleAlgo{ruled}
\begin{algorithm}[hbt!]
\caption{Smooth Residual Random Greedy (General Matroid)}\label{alg:matroid}
$S_0 \gets \emptyset$.\\
\For{$i=1,...,T$}{
    Let $M_i $ be a base of $\mathcal{M}/S_{i-1}$ maximizing $ \sum _{u\in M_i}\left[ f(S_{i-1}\cup \{u\})-f(S_{i-1})\right] $.\\
    \textbf{With probability $1-|S_{i-1}|/k$:\\}
     ~~~~~Let $u_i$ be a uniformly random element from $M_i$.\\
    ~~~~~$S_i \gets S_{i-1} \cup \{u_i\}$.\\
    \textbf{Otherwise} (with the complement probability of $|S_{i-1}|/k$)\\ 
    ~~~~~$S_i \gets S_{i-1}$.\\
}
Return $S_T$.
\end{algorithm}

We define how an optimal extension to the current solution evolves with respect to the iterations of Algorithm \ref{alg:matroid}.
To this end, we need to extend the definition of $O_S$ as follows:
$ O_S\triangleq \argmax_{A} \{ f(S\cup A):A\in \mathcal{I}_{M/S}\}$ for every $ S\in \mathcal{I}$, {\em i.e.}, $A$ maximizes $ f(S\cup A)$ among all subsets in $ \mathcal{N}\setminus S$ whose union with $S$ is independent in the original matroid.
Recalling that we assumed without loss of generality that $\mathcal{N}$ is padded with  dummy elements 
one can note that for every $ S\in \mathcal{I}$: $ O_S$ is a base of $ \mathcal{M}/ S$ and $ S\cup O_S$ is a base of $ \mathcal{M}$.
Moreover, $ |M_i|=k-|S_{i-1}|$ for every $i=1,\ldots,k$. 

Similarly to the analysis of Algorithm \ref{alg:partition} the following two lemmas establish a system of joint recursive formulas for $\mathbb{E}[f(S_i)]$ and $\mathbb{E}\left[f\left(O_{S_i}\cup S_{i}\right)\right]$:
Lemmas \ref{lemma:exp of s m} and \ref{lemma:exp of xor m} correspond to Lemmas \ref{lemma:exp of s} and \ref{lemma:exp of xor}, respectively.
Since both pairs of lemmas provide the exact same guarantee, in the proofs of Lemmas \ref{lemma:exp of s m} and \ref{lemma:exp of xor m} we only present the difference in the argument that is required.
In order to do that we use the following observation which gives the probability of an element of $M_i$ to be chosen by the algorithm and added to the solution.

\begin{observation}\label{matroid prob}
For every $i=1,\ldots,T$ and for every $u\in M_i$, assuming $ |M_i|\geq 1$ the probability that $u$ is chosen during the $i$\textsuperscript{th} iteration of the algorithm equals $1/k$.
\end{observation}
\begin{proof}
A uniform random element of $M_i$ is chosen with a probability of $1-|S_{i-1}|/k$.
Therefore, we get that every element $u\in M_i$ is chosen during the $i$\textsuperscript{th} iteration with a probability of: $$\left( 1-\frac{|S_{i-1}|}{k}\right)\cdot \frac{1}{|M_i|} = \frac{k-|S_{i-1}|}{k} \cdot \frac{1}{k-|S_{i-1}|} = \frac{1}{k} . $$
The above first equality follows since $ |M_i|=k-|S_{i-1}|$ for every $i=1,\ldots,k$.
\end{proof}

\begin{lemma}\label{lemma:exp of s m}
For every $i=1,\ldots,T$:
\begin{align}
\mathbb{E}\left[f(S_i)\right] - \mathbb{E}\left[f(S_{i-1})\right]\geq \frac{1}{k}\left[f\left(O_{ S_{i-1}}\cup S_{i-1}\right)-f(S_{i-1})\right]. \nonumber
\end{align}

\begin{proof}
The differences between the proof of Lemma \ref{lemma:exp of s} and the current lemma are the following two.
First, we employ Observation \ref{matroid prob} to infer the equality in (\ref{ineq new}).
Second, following the above discussion and the definition of $O_{S_{i-1}}$ for a general matroid we know that
$O_{S_{i-1}}$ is a base of $\mathcal{M}/S_{i-1}$.
Thus, $O_{S_{i-1}}$ is a possible choice of $M_i$ but not necessarily an optimal one according to the greedy rule.
Hence, we can conclude that the inequality in (\ref{ineq1}) also holds.
The rest of the proof now follows.
%
\end{proof}
\end{lemma}


\begin{lemma}\label{lemma:exp of xor m}
For every $i=1,\ldots,T$:
\begin{align}
\mathbb{E}\left[f\left(O_{ S_i}\cup S_i\right)\right] \geq \left( 1-\frac{2}{k}\right) \cdot \mathbb{E}\left[f\left(O_{S_{i-1}}\cup S_{i-1}\right)\right]+\frac{1}{k}\cdot\mathbb{E}\left[ f(S_{i-1})\right].
\end{align}
\end{lemma}
\begin{proof}
The differences between the proof of Lemma \ref{lemma:exp of xor} and the current lemma are the following two.
First, in order to infer the equality in (\ref{eq:1}) one needs
Observation \ref{matroid prob}.
Second, to infer inequality (\ref{Os1}) one needs an appropriate definition of $h$ for a general matroid.
Let $h:M_i \xrightarrow{} O_{S_{i-1}} $ be a bijection mapping every element $u \in M_i$ to an element of $O_{ S_{i-1}} $ in such a way that $O_{S_{i-1}} \setminus \{h(u)\} \cup \{u\}$ is a base of $\mathcal{M}/S_{i-1}$.
We note that the above definition of $h$ for a general matroid and every iteration $i$ reduces to the $h$ used in the proof of Lemma \ref{lemma:exp of xor} in the case that $\mathcal{M}$ is a partition matroid.
Thus, we can conclude that the inequality in (\ref{Os1}) also holds.
The rest of the proof now follows.
\end{proof}

\begin{lemma}\label{lem:SmoothMatroid}
Algorithm \ref{alg:matroid} achieves an approximation ratio of at least $0.27493$ for the problem of maximizing a non-monotone submodular function subject to a matroid independence constraint.
\end{lemma}
\begin{proof}
The same proof as for Lemma \ref{lemme:alg:partition}.
\end{proof}

\begin{proof}[Proof of Theorem \ref{thrm:RRG}]
Follows from Lemma \ref{lem:SmoothMatroid}. 
\end{proof}

\subsection{Comparison of Smoothed and Original Residual Random Greedy}\label{comparison}

First, let us note that if the original \RRG algorithm performs a fixed number $ \ell$ of iterations (where $ \ell\in \{ 1,2,\ldots,k\}$), then this is the same as the smoothed \RRG algorithm (Algorithm \ref{alg:partition}) performing a random number $T_{\ell}$ of iteration.
Let us now elaborate on the distribution of $ T_{\ell}$.
For every $ i\in\{ 0,1,\ldots,k-1\}$, let $Z_i$ be the random variable that equals the number of iterations Algorithm \ref{alg:partition} performs in order to decrease the number of parts of the partition the algorithm's solution intersects by one, given that $i$ parts in the partition are already intersected.
Note that $Z_i\sim Geom(1-\nicefrac{i}{k})$, for every $ i\in \{0,1,\ldots,k-1\}$ and all $Z_i$s are independent.
Therefore, in order to have exactly $\ell$ parts in the partition the algorithm's solution intersects (as is the case with the original \RRG algorithm after performing $\ell$ iterations), one needs to perform $Z_0+Z_1+\ldots+Z_{\ell-1}$ iterations of Algorithm \ref{alg:partition}.
Therefore, $ T_{\ell}=Z_0+\ldots + Z_{\ell-1}$.

Second, let us note that if the smoothed \RRG algorithm (Algorithm \ref{alg:partition}) performs a fixed number $T$ of iterations (where $ T\in \{ 1,2,\ldots\}$), then this is the same as the original \RRG algorithm performing a random number $ \ell_T$ of iterations.
Let us now elaborate on the distribution of $ \ell _T$.
The number of parts in the partition that intersect the solution generated by Algorithm \ref{alg:partition} after $ T$ iterations equals the smallest $ i\in \mathbb{N}$ such that: $ Z_0+Z_1+\ldots Z_i>T$.
Hence, $ \ell _T = \min \{ i\in \mathbb{N}:Z_0+\ldots + Z_i>T\}$.
One can note that by definition $ \ell _T\in \{ 1,2,\ldots,k\}$ since $Z_k\sim Geom(0)$ and thus $ Z_k=\infty$ with a probability of $1$.


In the context of a uniform random arrival order, following the above discussion, there is a distinction between the original \RRG algorithm and Algorithm \ref{alg:partition} (assuming the latter performs a fixed deterministic number of iterations).
The original \RRG algorithm is actually a deterministic greedy algorithm: once the next (random) item arrives it is {\em deterministically} assigned to the bidder with the highest marginal value, assuming this marginal is non-negative (otherwise the item is discarded).
However, if the smooth version (Algorithm \ref{alg:partition}) performs a fixed deterministic number $T$ of iterations then it is in fact a randomized algorithm.
Specifically, if Algorithm \ref{alg:partition} performs $T$ iterations, then this corresponds to examining a {\em random} number $\ell_T$ of the first items to arrive (and deterministically assigning each of these items to the bidder with highest marginal value at the moment of arrival, if such a bidder exists) and discarding the rest of the items regardless of their marginal values.

\newpage

\bibliography{lipics-v2021-sample-article}

\appendix

\section{Proof of Lemma \ref{lemma:both}}\label{app:lem_both}
\begin{proof}
We prove the lemma by induction on $i$.
For $i=0$ the lemma holds due to the following two.
First, $S_0=\emptyset$ and thus $ f(S_0)\geq 0$ from the non-negativity of $f$.
Moreover, the right hand side of (\ref{ineq::1}) also equals $0$ when $i=0$.
Hence, we can conclude that $ \mathbb{E}[f(S_0)]$ satisfies the required inequality for $i=0$.
Second, without loss of generality $ O_{S_0}\cup S_0=OPT$ since $S_0=\emptyset$ and thus $ f( O_{S_o}\cup S_0)=f(OPT)$.
Moreover, the right hand side of (\ref{ineq::2}) also equals $ f(OPT)$ when $i=0$.
Thus, we can conclude that $ \mathbb{E}[f(O_{S_0}\cup S_0)]$ satisfies the required inequality for $i=0$.

For $i>0$ we first focus on (\ref{ineq::1}):
\begin{align}
    \mathbb{E}[f(S_i)] \geq & \left( 1-\frac{1}{k}\right) \cdot \mathbb{E}[f(S_{i-1})]+ \frac{1}{k}\cdot \mathbb{E}[f(O_{S_{i-1}}\cup S_{i-1})] \label{induction_1}\\
    \geq & \left( 1-\frac{1}{k}\right) \frac{f(OPT)}{\sqrt{5}}\left[ \left( 1-\frac{a}{k}\right)^{i-1}-\left( 1-\frac{b}{k} \right)^{i-1}\right] + \label{induction_2}\\
    & \frac{1}{k}\frac{f(OPT)}{2\sqrt{5}}\left[ (\sqrt{5}-1)\left( 1-\frac{a}{k}\right)^{i-1}+(\sqrt{5}+1)\left( 1-\frac{b}{k}\right)^{i-1}\right] \nonumber\\
    = & \frac{f(OPT)}{\sqrt{5}}\left\{ \left( 1-\frac{a}{k} \right)^{i-1} \left[ 1-\frac{1}{k}+\frac{\sqrt{5}-1}{2k}\right]-\left( 1-\frac{b}{k}\right)^{i-1}\left[ 1-\frac{1}{k}-\frac{\sqrt{5}+1}{2k}\right]\right\} \label{induction_3} \\
    = & \frac{f(OPT)}{\sqrt{5}} \left( \left( 1-\frac{a}{k}\right)^{i}-\left( 1-\frac{b}{k}\right)^{i}\right). \label{induction_4}
\end{align}
Equality (\ref{induction_1}) follows from Lemma \ref{lemma:exp of s}.
We note that inequality (\ref{induction_2}) follows from the induction hypothesis for both $ \mathbb{E}[f(S_{i-1})]$ and $ \mathbb{E}[f(O_{S_{i-1}}\cup S_{i-1})]$.
Equality (\ref{induction_3}) follows from rearranging terms.
The last equality (\ref{induction_4}) holds since $ a=(3-\sqrt{5})/2$ and $ b=(3+\sqrt{5})/2$.

Let us now focus on (\ref{ineq::2}):
\begin{align}
    & \mathbb{E}[f(O_{S_i}\cup S_i)] \geq \left( 1-\frac{2}{k}\right) \cdot \mathbb{E}[f(O_{S_{i-1}}\cup S_{i-1})]+\frac{1}{k}\cdot \mathbb{E}[f(S_{i-1})] \label{induction_5}\\
    \geq & \left( 1-\frac{2}{k}\right) \frac{f(OPT)}{2\sqrt{5}} \left[ (\sqrt{5}-1)\left( 1-\frac{a}{k}\right)^{i-1}+(\sqrt{5}+1)\left( 1-\frac{b}{k}\right)^{i-1}\right] +\label{induction_6}\\
    & \frac{1}{k}\frac{f(OPT)}{\sqrt{5}}\left[ \left( 1-\frac{a}{k}\right)^{i-1} - \left(1-\frac{b}{k} \right)^{i-1}\right] \nonumber \\
    = & \frac{f(OPT)}{2\sqrt{5}}\left\{ \left( 1-\frac{a}{k}\right)^{i-1}\left[ \left( 1-\frac{2}{k}\right)(\sqrt{5}-1)+\frac{2}{k}\right] + \left( 1-\frac{b}{k}\right)^{i-1}\left[ \left( 1-\frac{2}{k}\right)(\sqrt{5}+1)-\frac{2}{k}\right]\right\} \label{induction_7}\\
    = & \frac{f(OPT)}{2\sqrt{5}}\left( (\sqrt{5}-1)\left( 1-\frac{a}{k}\right)^i+(\sqrt{5}+1)\left( 1-\frac{b}{k}\right)^i\right).\label{induction_8}
\end{align}
Equality (\ref{induction_5}) follows from Lemma \ref{lemma:exp of xor}.
We note that inequality (\ref{induction_6}) follows from the induction hypothesis for both $ \mathbb{E}[f(O_{S_{i-1}}\cup S_{i-1})]$ and $ \mathbb{E}[f(S_{i-1})]$.
Equality (\ref{induction_7}) follows from rearranging terms.
The last equality (\ref{induction_8}) holds since $ (1-2/k)(\sqrt{5}-1)+2/k=(\sqrt{5}-1)(1-a/k)$ (recalling that $ a=(3-\sqrt{5})/2$) and $ (1-2/k)(\sqrt{5}+1)-2/k=(\sqrt{5}+1)(1-b/k)$ (recalling that $ b=(3+\sqrt{5})/2$).
\end{proof}

\end{document}